\documentclass[lettersize,journal]{IEEEtran}
\usepackage{amsmath,amsfonts}
\usepackage{lipsum}
\usepackage{algorithmic}
\usepackage{algorithm}
\usepackage{array}
\usepackage{subfigure}
\usepackage{textcomp}
\usepackage{stfloats}
\usepackage{url}
\usepackage{verbatim}
\usepackage{graphicx}
\usepackage{cite}
\usepackage{xcolor}
\usepackage{enumerate}

\begin{document}

\title{How to Extend 3D GBSM to RIS Cascade Channel with Non-ideal Phase Modulation?}

\author{Huiwen Gong, Jianhua Zhang, Yuxiang Zhang, Zhengfu Zhou, and Guangyi Liu
        % <-this % stops a space

\thanks{Manuscript received xxx; revised xxx; accepted xxx. Date of publication xxx; date of current version xxx. This work was supported in part by the National Science Fund for Distinguished Young Scholars under Grant 61925102; in part by the National Natural Science Foundation of China under Grant 92167202; in part the National Key Research and Development Program of China under Grant 2020YFB1805002; and in part by the BUPT-CMCC Joint Innovation Center. The associate editor coordinating the review of this article and approving it for publication was xxx. (Corresponding author: xxx.)

Huiwen Gong, Jianhua Zhang, Yuxiang Zhang, and Zhengfu Zhou are with the State Key Laboratory of Networking and Switching Technology, Beijing University of Posts and Telecommunications, Beijing 100876, China (e-mail: birdsplan@bupt.edu.cn; jhzhang@bupt.edu.cn; zhangyx@bupt.edu.cn; zhengfu@bupt.edu.cn;). Guangyi Liu is with the Future Research Laboratory, China Mobile Research Institute, Beijing 100053, China (e-mail: liuguangyi@chinamobile.com;).

Digital Object Identifier xxx.}
}

\markboth{Journal of \LaTeX\ Class Files,~Vol.~14, No.~8, August~2021}%
{Shell \MakeLowercase{\textit{et al.}}: A Sample Article Using IEEEtran.cls for IEEE Journals}


\maketitle

\begin{abstract}

Reconfigurable intelligent surface (RIS) is envisioned as a promising technology for next-generation wireless communications.
Its deployment introduces a RIS cascade link between the transmitter (Tx) and receiver (Rx), which makes its channel model significantly different from the Tx-Rx direct link.
In this letter, a RIS cascade channel modeling method based on a 3D geometry-based stochastic model (GBSM) is proposed.
The model follows a 3GPP standardized modeling framework and extends the traditional Tx-Rx channel to Tx-RIS-Rx cascade channel.
In the modeling process, we consider the non-ideal phase modulation of the RIS element, so as to accurately characterize the dependence of its phase modulation on the incoming wave angle.
The differences between the proposed cascade channel model and the channel model with ideal phase modulation are investigated.
The simulation results show that the proposed model can better reflect the dependence of RIS on angle and polarization. 

\end{abstract}


\section{Introduction}
As one of the key technologies of 6G, reconfigurable intelligent surface (RIS) has broad application prospects in enhancing the quality of communication links and increasing the communication coverage area.
Similar to the 3D MIMO technology in 5G \cite{3dmimo}, RIS is also a two-dimensional plane. The difference is that RIS compose of a large number of periodically arranged metamaterial elements \cite{renzo1}. Each element can independently change the phase of the arriving signal to achieve artificial control of the reflected signal.

RIS is often deployed between a transmitter (Tx) and a receiver (Rx), which means that a Tx-RIS-Rx cascade link will be introduced in addition to the conventional Tx-Rx link. 
To analyze and optimize the RIS-assisted wireless communication system, an accurate and easy-to-implement RIS cascade channel model is required.
Some works consider the case that there is no scatterer in the channel whether at the transceiver or RIS \cite{twc,twc2,renzo2}.
In \cite{twc}, the authors propose a free-space propagation channel model of the Tx-RIS-Rx cascade link, and verify the proposed model by conducting channel measurements in a microwave anechoic chamber.
The authors in \cite{twc2} consider the direct link between transceivers, and propose a two-path propagation model. Under the assumption of line-of-sight, a RIS-assisted communication channel model based on mutual impedances is proposed in \cite{renzo2}.
Some other works consider more practical channel conditions, which assume that multipath fading exists in RIS channel \cite{vincent,wqq,heruisi1}.
In \cite{vincent} and \cite{wqq}, a quasi-static flat fading model is considered in both Tx-RIS and RIS-Rx channels, and the RIS cascade channel is the multiplication of the two sub-channels and the phase modulation matrix.
While in \cite{heruisi1}, the authors consider a dominant path from RIS to Rx, so Rician fading is used to represent the RIS-Rx channel.

 %下一步：GBSM，信道之间不独立（调相与径有关）
The above works make some specific assumptions about the channel, such as RIS working in free space, or in a flat fading channel.
However, channel measurement shows that the channel situation is complex \cite{liyi,jsac}, and the scatterers present different distributions. Since the geometry-based stochastic model (GBSM) has natural advantages in describing the characteristics of scatterers and has been adopted by ITU and 3GPP as a standard channel modeling method, it is expected to apply to the RIS channel.
Besides, the above works adopt an ideal phase modulation model to describe RIS itself, in which RIS can achieve a constant phase modulation for signals at any angle of arrival.
However, some works show that the phase modulation of RIS is angle-dependent and non-ideal \cite{zhengfu1,twcangle}. In other words, different angles of the incident path will affect the phase modulation of RIS, which leads to a deviation between the actual phase shift of RIS and the target phase shift.

Therefore, this letter proposes a GBSM principle based RIS cascade channel model with non-ideal phase modulation.
This model follows a 3GPP standardized modeling framework and considers an angle-dependent phase modulation of RIS.
The influence of angle on phase modulation and the scattering gain of RIS are both reflected in the equivalent radiation pattern of RIS.
Then the specific implementation of the RIS cascade channel model is provided based on the extended 3GPP channel model framework.
Based on the proposed channel model, the impact of non-ideal phase modulation on the RIS cascade channel is investigated.

\section{The proposed RIS cascade channel model}
\begin{figure}[htbp]
\centerline{\includegraphics[width=8.5cm]{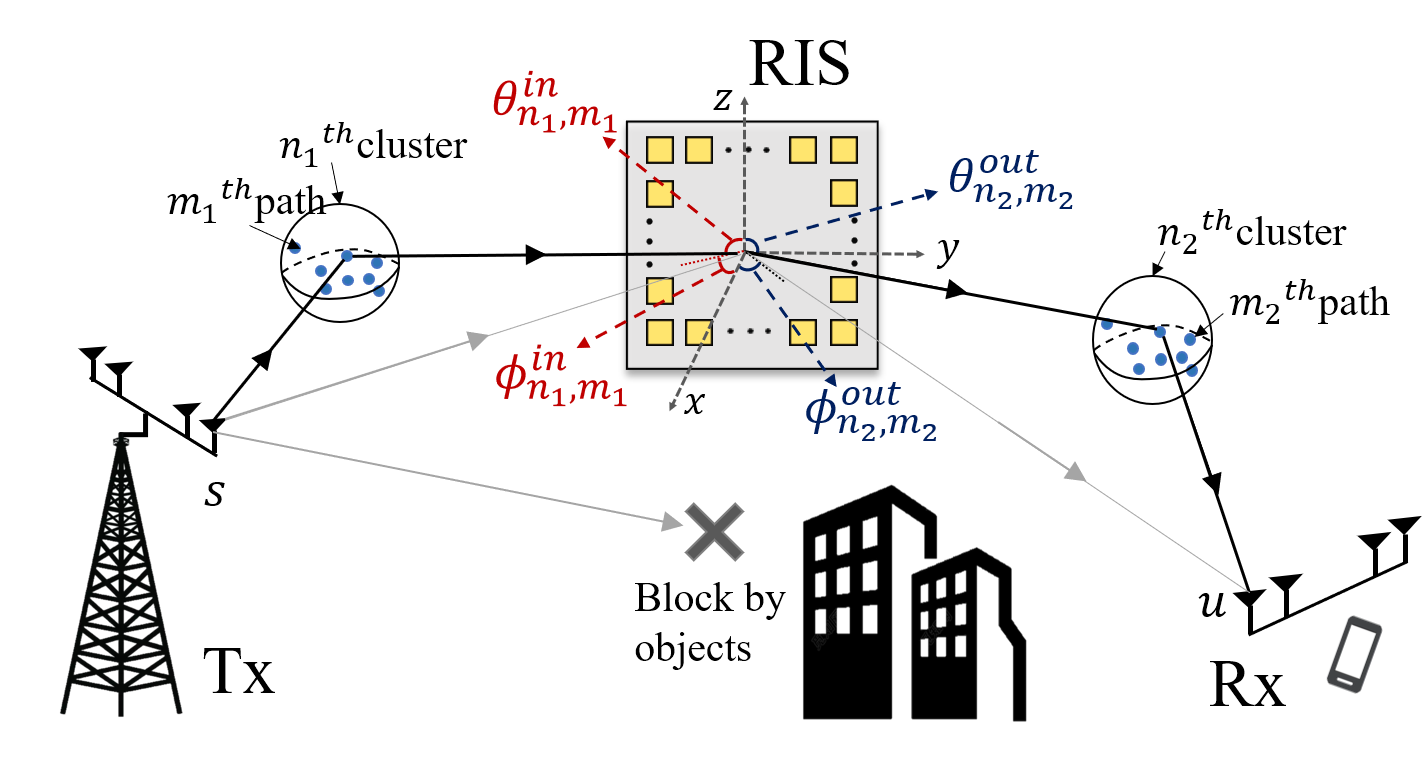}}
\caption{The system model of RIS-assisted MIMO communication. There are $N_{1}$ clusters and $M_{1}$ paths in the Tx-RIS channel, $N_{2}$ clusters and $M_{2}$ paths in the RIS-Rx channel. The direct link between Tx and Rx is blocked by objects.}
\label{fig0}
\end{figure}

In this section, a framework of channel modeling for the RIS cascade channel is proposed.
Consider a RIS-assisted MIMO communication scenario illustrated in Fig. \ref{fig0}.

\subsection{RIS cascade channel with ideal phase modulation}

Under the assumption that the phase modulation of RIS is ideal and independent from the path angle in the Tx-RIS and RIS-Rx sub-channels, the RIS cascade channel between  the $s$-th transmitter and the $u$-th receiver $h_{u,s}^{ris}$ can be expressed as the multiplication of two independent sub-channels and the phase modulation matrix, written as
\begin{equation}
h_{u,s}^{ris} = \sum\limits_{n_{2},m_{2}} h_{n_{2},m_{2}}^{u,ris}
\mathbf{\Theta} \sum\limits_{n_{1},m_{1}}h_{n_{1},m_{1}}^{ris,s}.    
\end{equation}
where $h_{n_{2},m_{2}}^{u,ris}$ and $h_{n_{1},m_{1}}^{ris,s}$ respectively represent the channel coefficient for $(n_{2},m_{2})^{th}$ path in RIS-Rx sub-channel and for $(n_{1},m_{1})^{th}$ path in Tx-RIS sub-channel. $\mathbf{\Theta}$ represents the phase modulation coefficients matrix of RIS, written as $\boldsymbol{\Theta}=\operatorname{diag}\left(e^{j \alpha_{1}}, \cdots, e^{j \alpha_{N}}\right)$. $\alpha_{n}$ represents the phase shift of $n^{th}$ RIS element.

\begin{figure*}[hb]
\begin{equation}
\begin{aligned}
 h_{u,s}^{ris}(t,\tau)  = 
&\sum_{n_{1},m_{1}}^{N_{1},M_{1}}\sum_{n_{2},m_{2}}^{N_{2},M_{2}}\sqrt{P_{n_{1},m_{1}}P_{n_{2},m_{2}}}\left[\begin{array}{c}
        F_{r x, u}^{v}\left(\theta_{n_{2},m_{2}}^{r x}, \phi_{n_{2},m_{2}}^{r x}\right) \\
        F_{r x, u}^{h}\left(\theta_{n_{2},m_{2}}^{r x}, \phi_{n_{2},m_{2}}^{r x}\right)
    \end{array}\right]^{\mathrm{T}} \cdot
\left[\setlength{\arraycolsep}{0.4pt}\begin{array}{cc}
\exp \left(j \Phi_{n_{2},m_{2}}^{vv}\right) & \sqrt{\kappa_{n_{2},m_{2}}^{-1}} \exp \left(j \Phi_{n_{2},m_{2}}^{vh}\right) \\
\sqrt{\kappa_{n_{2},m_{2}}^{-1}} \exp \left(j \Phi_{n_{2},m_{2}}^{hv}\right) & \exp \left(j \Phi_{n_{2},m_{2}}^{hh}\right)
\end{array}\right]\\
&\cdot\left[\begin{array}{cc}
F_{ris}^{vv}(\phi_{n_{1},m_{1}}^{in}, \theta_{n_{1},m_{1}}^{in}, \phi_{n_{2},m_{2}}^{out}, \theta_{n_{2},m_{2}}^{out})  &
F_{ris}^{vh}(\phi_{n_{1},m_{1}}^{in}, \theta_{n_{1},m_{1}}^{in}, \phi_{n_{2},m_{2}}^{out}, \theta_{n_{2},m_{2}}^{out}) \\
F_{ris}^{hv}(\phi_{n_{1},m_{1}}^{in}, \theta_{n_{1},m_{1}}^{in}, \phi_{n_{2},m_{2}}^{out}, \theta_{n_{2},m_{2}}^{out}) &
F_{ris}^{hh}(\phi_{n_{1},m_{1}}^{in}, \theta_{n_{1},m_{1}}^{in}, \phi_{n_{2},m_{2}}^{out}, \theta_{n_{2},m_{2}}^{out})
\end{array}\right]\\&
\cdot\left[\begin{array}{cc}
\exp \left(j \Phi_{n_{1},m_{1}}^{vv}\right) & \sqrt{\kappa_{n_{1},m_{1}}^{-1}} \exp \left(j \Phi_{n_{1},m_{1}}^{vh}\right) \\
\sqrt{\kappa_{n_{1},m_{1}}^{-1}} \exp \left(j \Phi_{n_{1},m_{1}}^{hv}\right) & \exp \left(j \Phi_{n_{1},m_{1}}^{hh}\right)
\end{array}\right]
\cdot
\left[\begin{array}{c}
        F_{tx, s}^{v}\left(\theta_{n_{1},m_{1}}^{t x}, \phi_{n_{1},m_{1}}^{t x}\right) \\
        F_{tx, s}^{h}\left(\theta_{n_{1},m_{1}}^{t x}, \phi_{n_{1},m_{1}}^{t x}\right)
    \end{array}\right]\\
&\cdot\exp \left(j \frac{2\pi}{\lambda }\left(\boldsymbol{\rm r}_{n_{2},m_{2}}^{rx} 
\cdot \boldsymbol{\rm d}_{u}^{rx}+\boldsymbol{\rm r}_{n_{1},m_{1}}^{tx} \cdot \boldsymbol{\rm d}_{s}^{tx}\right)\right)
\cdot \exp (j 2\pi f_{n_{2},m_{2}} t)
\cdot\delta{(\tau-\tau_{n_{1},m_{1}}-\tau_{n_{2},m_{2}})} .
\end{aligned}
\label{htotal}
\end{equation}
\end{figure*}

\subsection{RIS cascade channel with non-ideal phase modulation}
Here, we attempt to extend the 3GPP GBSM method and give a representation of the RIS cascade channel under non-ideal phase modulation.
As mentioned in the previous research, RIS phase modulation has some dependency on the path angle \cite{zhengfu1,twcangle}, which can be reflected in the radiation pattern of RIS.
Considering the above properties, the channel impulse response (CIR) of the RIS-assisted communication link between $s$ and $u$ can be expressed as \eqref{htotal}.
The RIS cascade channel model is defined in spherical coordinates, where
\begin{itemize}
    \item $(\cdot)^{T}$ stands for matrix transposition.
    \item $\lambda$ is the wavelength of the carrier frequency.
    \item $n_{i},m_{i}$, $i \in \{1,2\}$ indicate the identifier of cluster and path in Tx-RIS and RIS-Rx sub-channels.
    \item $P_{n_{1},m_{1}}$ and $P_{n_{2},m_{2}}$ are the power of the corresponding path.
    \item $\phi_{n_{1},m_{1}}^{in}, \theta_{n_{1},m_{1}}^{in},\theta_{n_{1},m_{1}}^{t x}, \phi_{n_{1},m_{1}}^{t x}$ denote the angle of $(n_{1},m_{1})^{th}$ path in Tx-RIS sub-channel, which are the azimuth angle of arrival (AoA), the zenith angle of arrival (ZoA), the azimuth angle of departure (AoD), and the zenith angle of departure (ZoD) respectively. Similarly, $\phi_{n_{2},m_{2}}^{rx}, \theta_{n_{2},m_{2}}^{rx},\theta_{n_{2},m_{2}}^{out}, \phi_{n_{2},m_{2}}^{out}$ are the AoA, ZoA, AoD, ZoD of the $(n_{2},m_{2})^{th}$ path in RIS-Rx sub-channel.
    \item $\boldsymbol{\rm r}_{n_{2},m_{2}}^{rx}$ and $\boldsymbol{\rm r}_{n_{1},m_{1}}^{tx}$ are the unit direction vector of corresponding path at receiver and transmitter, taking $\boldsymbol{\rm r}_{n_{1},m_{1}}^{tx}$ as an example, it can be expressed as
    \begin{equation}
        \mathbf{r}^{tx}_{n_{1},m_{1}}=\left[\begin{array}{c}
        \sin \theta_{n_{1},m_{1}}^{tx} \cos \phi_{n_{1},m_{1}}^{tx} \\
        \sin \theta_{n_{1},m_{1}}^{tx} \sin \phi_{n_{1},m_{1}}^{tx} \\
        \cos \theta_{n_{1},m_{1}}^{tx}
        \end{array}\right].
        \label{raydirtx}
    \end{equation}
    \item $\boldsymbol{\rm d}_{u}^{rx}$ and $\boldsymbol{\rm d}_{s}^{tx}$ denote the location vectors of antenna $u$ and $s$.
    \item $p_{1},p_{2} \in \{v,h\}$ denote the vertical and horizontal polarization direction.
    \item $F_{r x, u}^{v}, F_{r x, u}^{h}, F_{tx, s}^{v}$ and $F_{tx, s}^{h}$ indicate the radiation pattern of antenna $u$ at Rx and $s$ at Tx, in the $v$ and $h$ polarization, respectively.
    \item $F_{ris}^{p_{1}p_{2}}$ denotes the radiation pattern of entire RIS. $p_{1}p_{2}$ indicates that RIS has different effects on the cluster or path incident in the $p_{1}$ polarization direction and emitted in the $p_{2}$ polarization direction.
    \item $\Phi_{n_{i},m_{i}}^{p_{1}p_{2}}$, $i \in\{1,2\}$ is the random phase of the $(n_{i},m_{i})^{th}$ path that depart in $p_{1}$ direction and arrive in $p_{2}$ direction. And $\kappa_{n_{i},m_{i}}$, $i \in\{1,2\}$ is the cross polarization power ratio (XPR) of the corresponding path.
    \item $f_{n_{2},m_{2}}$ is the Doppler shift of the $(n_{2},m_{2})^{th}$ path.
    \item $\tau_{n_{i},m_{i}}$, $i \in\{1,2\}$ denotes the delay of the corresponding path.
\end{itemize}

Compared with the traditional 3D MIMO channel model defined by the 3GPP standard \cite{38901}, the proposed RIS cascade channel model includes Tx-RIS and RIS-Rx two sub-channels, and there will be two sets of clusters and paths parameters accordingly.
The effect of RIS on clusters and paths is contained in the equivalent RIS radiation pattern $F_{ris}^{p_{1}p_{2}}$.
In the next subsection, we will utilize the surface equivalence theorem to obtain $F_{ris}^{p_{1}p_{2}}$.

\subsection{Radiation pattern of RIS}
\label{pattern}
RIS can affect the signals in the channel, which includes two aspects, one is the phase modulation effect of RIS, and the other is the scattering gain brought by the physical aperture of RIS. 
We will embody these properties in the radiation pattern of RIS in this subsection.

Assuming RIS is positioned on the $yoz$-plane, and the $(n_{1},m_{1})^{th}$ path incident on RIS element at $(x, y)^{th}$ position shown in Fig. \ref{fig0}.
Take $\mathbf{E}_{i}$ as the incident electric field.
$R_{x,y}$ represents the preset phase modulation on the $(x,y)^{th}$ RIS element. Take $R_{x,y}^{real}$ as the actual phase modulation of RIS element for $(n_{1},m_{1})^{th}$ path. As mentioned in subsection A, the phase modulation of RIS is non-ideal, which means $R_{x,y}^{real}$ does not coincide with $R_{x,y}$, but is influenced by $\theta_{n_{1},m_{1}}^{in}$.
Considering the impedance-type boundary conditions, $R_{x,y}^{real}$ can be calculated as
\begin{equation}
R_{x,y}^{real} = \frac{(1+R_{x,y})\eta -({1-R_{x,y}})\eta_{e}}{(1+R_{x,y})\eta +({1-R_{x,y}})\eta_{e}},
\label{rele}
\end{equation}
where $\eta_{e}$ donates the equivalent wave impedance. 
According to \cite{zhengfu1}, when the incident path is in vertical polarization, $\eta_{e}=\frac{\eta}{\cos\theta_{n_{1},m_{1}}^{in}}$, while in horizontal polarization, $\eta_{e}=\eta{\cos\theta_{n_{1},m_{1}}^{in}}$. And $\eta$ denotes the vacuum impedance.

To calculate the RIS radiation pattern, we first calculate the equivalent current $\mathbf{J}_{s}$ and equivalent magnetic current $\mathbf{M}_{s}$ of the RIS element according to equivalence theorem
\begin{equation}
\begin{aligned}
\mathbf{J}_{s} =&\hat{\boldsymbol{n}} \times\left(1-R_{x,y}^{real}\right)\mathbf{H}_{i}, \\
\mathbf{M}_{s} =&-\hat{\boldsymbol{n}} \times\left(1+R_{x,y}^{real}\right)\mathbf{E}_{i},
\end{aligned}
\label{JM1}
\end{equation}
where $\mathbf{H}_{i}$ is the magnetic field, and can be written as $\mathbf{H}_{i}=\frac{1}{\eta}\mathbf{e}_{n_{1},m_{1}}\times \mathbf{E}_{i}$. $\mathbf{e}_{n_{1},m_{1}}$ represents the unit direction vector of $(n_{1},m_{1})^{th}$ path.
Assuming a observing point $o$ in the direction of $\phi_{n2,m2}^{out}, \theta_{n2,m2}^{out}$, the magnetic vector potential $\mathbf{A}$ and electric vector potential $\mathbf{F}$ at observing point $o$ can be calculated as
\begin{equation}
\begin{aligned}
\mathbf{A} &=\iint_{S} \mu \mathbf{J}_{s} \frac{e^{-j \frac{2 \pi}{\lambda} R}}{4 \pi R} d s, \\
\mathbf{F} &= \iint_{S} \varepsilon \mathbf{M}_{s} \frac{e^{-j \frac{2 \pi}{\lambda} R}}{4 \pi R} d s,
\end{aligned}
\label{AF}
\end{equation}
where $S$ represents the area of the RIS element, $\mu$ and $\varepsilon$ denote the permeability and the permittivity in vacuum. $R$ is the distance from any point on RIS element to point $o$.
Scattering field $\mathbf{E}_{s}$ at point $o$ is
\begin{equation}
    \mathbf{E}_{s}=-j\omega\mathbf{A}-\frac{1}{\varepsilon} \nabla \times \mathbf{F},
\end{equation}
where $\varepsilon$ denotes permittivity. By removing the effect of distance of the observing point, the radiation pattern of the RIS element can be obtained as
\begin{equation}
f_{ris}^{p_{1}p_{2}}(\phi_{n_{1},m_{1}}^{in},\theta_{n_{1},m_{1}}^{in}, \phi_{n_{2},m_{2}}^{out}, \theta_{n_{2},m_{2}}^{out})=\frac{4\pi}{\lambda}\frac{r}{e^{-j \frac{2 \pi}{\lambda} r}}\frac{(\mathbf{E}_{s})_{p_{2}}}{(\mathbf{E}_{i})_{p_{1}}},
\label{V-in-element}
\end{equation}
where $r$ represents the distance between the center of the RIS element and point $o$.
$\frac{4\pi}{\lambda}$ is caused by the conversion between field strength and amplitude of path.
$(\mathbf{E}_{i})_{p_{2}}$ means the $p_{2}$ polarization component of $\mathbf{E}_{i}$.
$f_{ris}^{p_{1}p_{2}}$ represents the combination of different polarization of the incident and reflected waves, for example, $f_{ris}^{vh}$ means the horizontal polarization component of the reflected wave excited by the vertical polarization component of the incident wave at the RIS element.

The radiation pattern of the entire RIS is calculated as
\begin{equation}
\begin{aligned}
 F_{ris}^{p_{1}p_{2}}(\phi_{n_{1},m_{1}}^{in},&\theta_{n_{1},m_{1}}^{in}, \phi_{n_{2},m_{2}}^{out}, \theta_{n_{2},m_{2}}^{out})  = \\
&\sum_{x,y}^{X,Y} f_{ris}^{p_{1}p_{2}}(\phi_{n_{1},m_{1}}^{in}, \theta_{n_{1},m_{1}}^{in}, \phi_{n_{2},m_{2}}^{out}, \theta_{n_{2},m_{2}}^{out})\\
&\cdot  e^{\frac{2\pi}{\lambda }(\mathbf{r}^{in}_{n_{1},m_{1}} \cdot \mathbf{d}_{x,y})}e^{\frac{2\pi}{\lambda }(\mathbf{r}^{out}_{n_{2},m_{2}} \cdot \mathbf{d}_{x,y} )},
\end{aligned}
\label{panel}
\end{equation} 
where $X, Y$ is the number of rows and columns of RIS, and $\mathbf{r}^{in}_{n_{1},m_{1}}$ and $\mathbf{r}^{out}_{n_{2},m_{2}}$ denote the direction vector of the incident wave and the outgoing wave respectively. The calculation formula is similar to \eqref{raydirtx}.
$\mathbf{d}_{x,y}$ is the position vector of the $(x,y)^{th}$ RIS element in the panel. We take the center of the RIS board as the reference point, $\mathbf{d}_{x,y}$ can be expressed as
\begin{equation}
\mathbf{d}_{x,y}=\left[\begin{array}{c}
(x-\frac{1+X }{2})  \\
(y- \frac{1+Y }{2})\\
0
\end{array}\right]d,
\end{equation}
where $d$ is the interval between RIS elements.

\subsection{RIS cascade channel modeling implementation}
\begin{figure}[htbp]
\centerline{\includegraphics[width=9cm]{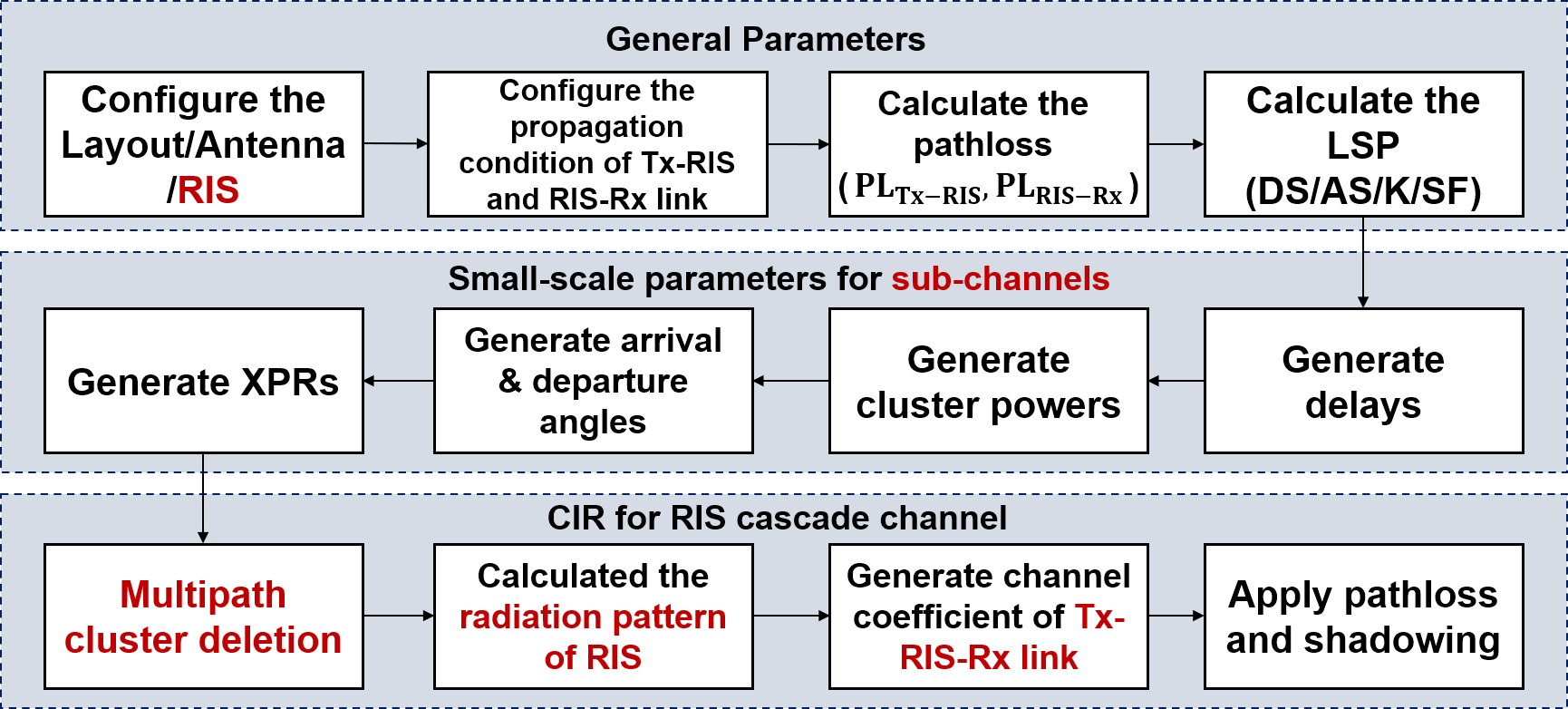}}
\caption{RIS cascade channel model implementation framework.}
\label{flow}
\end{figure}

Based on the proposed RIS channel model, we extend the current 3GPP-like 3D GBSM implementation framework as Fig. \ref{flow}.

This framework follows the procedure of the channel modeling in 3GPP \cite{38901}, and it adds steps such as RIS parameter configuration, sub-channels parameter generation, multipath cluster deletion, RIS radiation pattern calculation, and RIS cascade channel generation.
In summary, the modeling framework can be divided into three parts:

\emph{1) General Parameters:} Determine communication scenarios, generate general parameters of TX-RIS channel and RIS-Rx channel, and calculate the large-scale parameters (LSPs) and path loss of these two sub-channels.
The LSPs contain delay spread (DS), azimuth spread of arrival (ASA), zenith spread of arrival (ZSA), azimuth spread of departure (ASD), zenith spread of departure (ZSD), shadow fading (SF), and Rician K factor.
The correlation between these parameters is introduced through a cross-correlation matrix. The path loss of both sub-channels ($PL_{tx-ris}, PL_{ris-rx}$) is calculated according to \cite{38901}.

\emph{2) Small-scale Parameters:} Generate the small-scale parameters (SSPs) of the two sub-channels, which contain delays, powers, angles, and XPRs of clusters and paths.
The delay ($\tau_{n_{1},m_{1}},\tau_{n_{2},m_{2}}$) and power ($P_{n_{1},m_{1}}, P_{n_{2},m_{2}}$) of clusters are randomly generated by using exponential power delay distribution. 
Obtaining the ZoA, AoA, ZoD, and AoD of the relevant statistical distribution on both sub-channels through equal power sampling, where the zenith angle follows the Laplacian distribution, and the azimuth angle follows the Wrapped Gaussian distribution. 
And the XPRs ($\kappa_{n_{1},m_{1}},\kappa_{n_{2},m_{2}}$) for each path are generated by a log-normal distribution. The detailed generation process is explained in section 7.5 in \cite{38901}.
The initial phase of paths in sub-channel $\{\phi_{n_{1},m_{1}}^{vv},\phi_{n_{1},m_{1}}^{vh},\phi_{n_{1},m_{1}}^{hv},\phi_{n_{1},m_{1}}^{hh}\}$ should be generated, they follow a uniform distribution as $\mathcal{U}(0,2\pi)$.

\emph{3) Channel Impulse Response:} 
Considering the complexity of RIS radiation pattern calculation, we first set the power threshold and delete clusters and paths with lower power to reduce the complexity of the model. Then calculate the radiation pattern of RIS ($\mathbf{F}_{ris}^{p_{1}p_{2}}$) according to the above method. Finally, the RIS cascade channel ($h_{u,s}^{ris}$) is generated based on multipath combining. In the far-field case, the path loss of the RIS cascade channel can be expressed as the product of the path loss of two sub-channels.

Based on the above framework and principle, the RIS channel simulation platform has been developed and released in \cite{platform}.

\section{Numerical analysis}

In this section, simulations have been set up to validate the rationality of the proposed model. 

Here, we define three operating regimes of RIS: \emph{continuous anomalous reflection}, \emph{1bit anomalous reflection}, and \emph{specular reflection}.
The phase modulation of $(x,y)^{th}$ RIS element corresponding to the different regimes are
\begin{small}
\begin{equation}
\begin{array}{ll}
 R_{x,y}=e^{-j\frac{2\pi}{\lambda}(\mathbf{r}^{in}\cdot \mathbf{d}_{x,y}+\mathbf{r}^{out}\cdot\mathbf{d}_{x,y})}, &\begin{small}\text{Optimal anomalous reflection}\end{small}. \\
 R_{x,y}=\left\{\begin{array}{lc}
1 &  Re(R_{x,y})>0\\
e^{-j\pi} & Re(R_{x,y})<0 \\
\end{array}\right.,&\begin{small}\text{1 bit anomalous reflection}\end{small}. \\
R_{x,y}=1,   &\begin{small}\text{Specular reflection} \end{small}.
\end{array}  
\end{equation}    
\end{small}

\begin{figure}[htbp]
\centering
\subfigure[Radiation pattern of RIS with non-ideal phase modulation and ideal phase modulation.]{
\includegraphics[width=8.5cm]{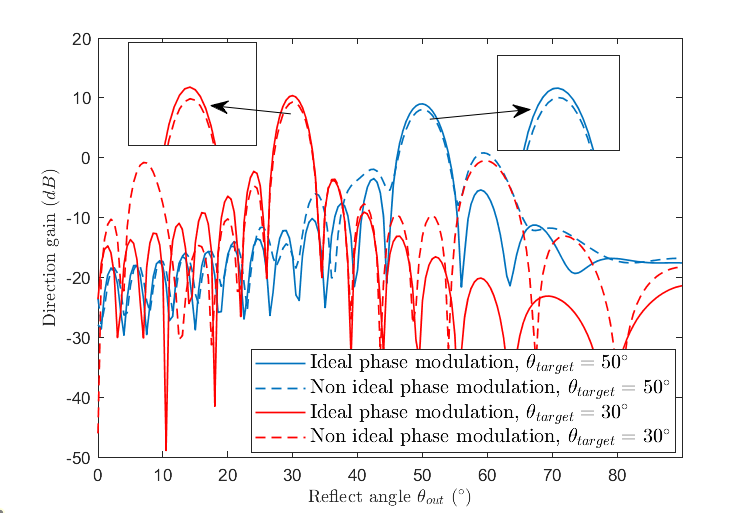}
\label{fig1-1}
}
\subfigure[Radiation pattern of RIS with different polarization.]{
\includegraphics[width=8.5cm]{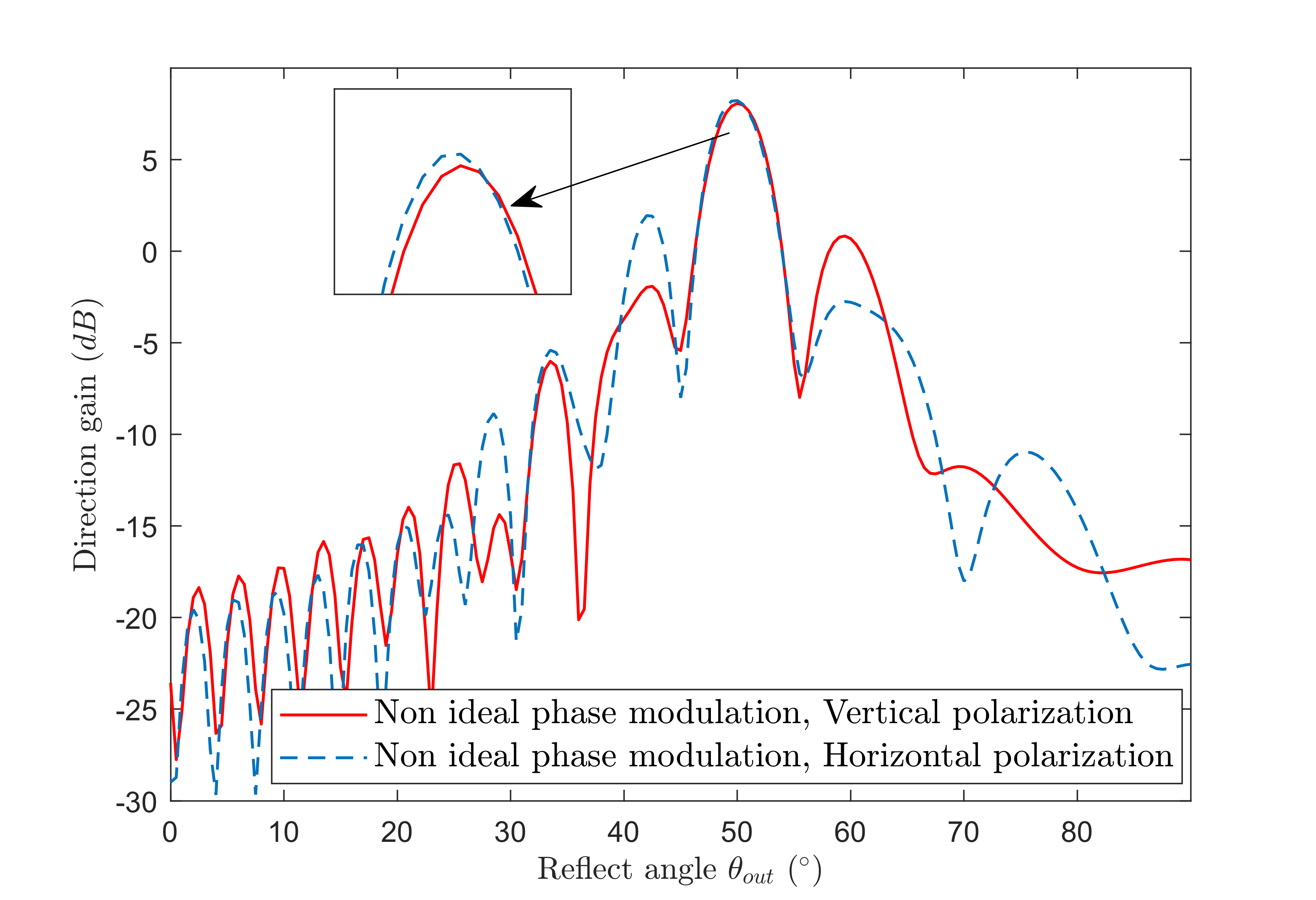}
\label{fig1-2}
}
\caption{The radiation pattern of RIS with different phase modulation and polarization.}\label{fig1}
\end{figure}

The radiation pattern of RIS with non-ideal phase modulation and ideal phase modulation is shown in Fig. \ref{fig1-1}, and the radiation pattern with different polarization components is shown in Fig. \ref{fig1-2}.
Suppose a $32\times32$ element RIS with a working frequency of 6 GHz, the interval between each element sets as half wavelength.
One path arrives at RIS with ZoA of $60^{\circ}$. The operating regime is set to optimal anomalous reflection. With this operation strategy, each element of the RIS is optimal phase modulated so that the target beam is oriented towards ZoD of $\theta_{target}$.
Fig. \ref{fig1-1} shows that compared to the RIS model with ideal phase modulation, the non-ideal phase modulation model causes about 1 dB of gain attenuation in the direction of the target beam.
Fig. \ref{fig1-2} shows that the radiation pattern of RIS is also affected by the polarization of the incident path.

\begin{figure}[htbp]
\centerline{\includegraphics[width=8.5cm]{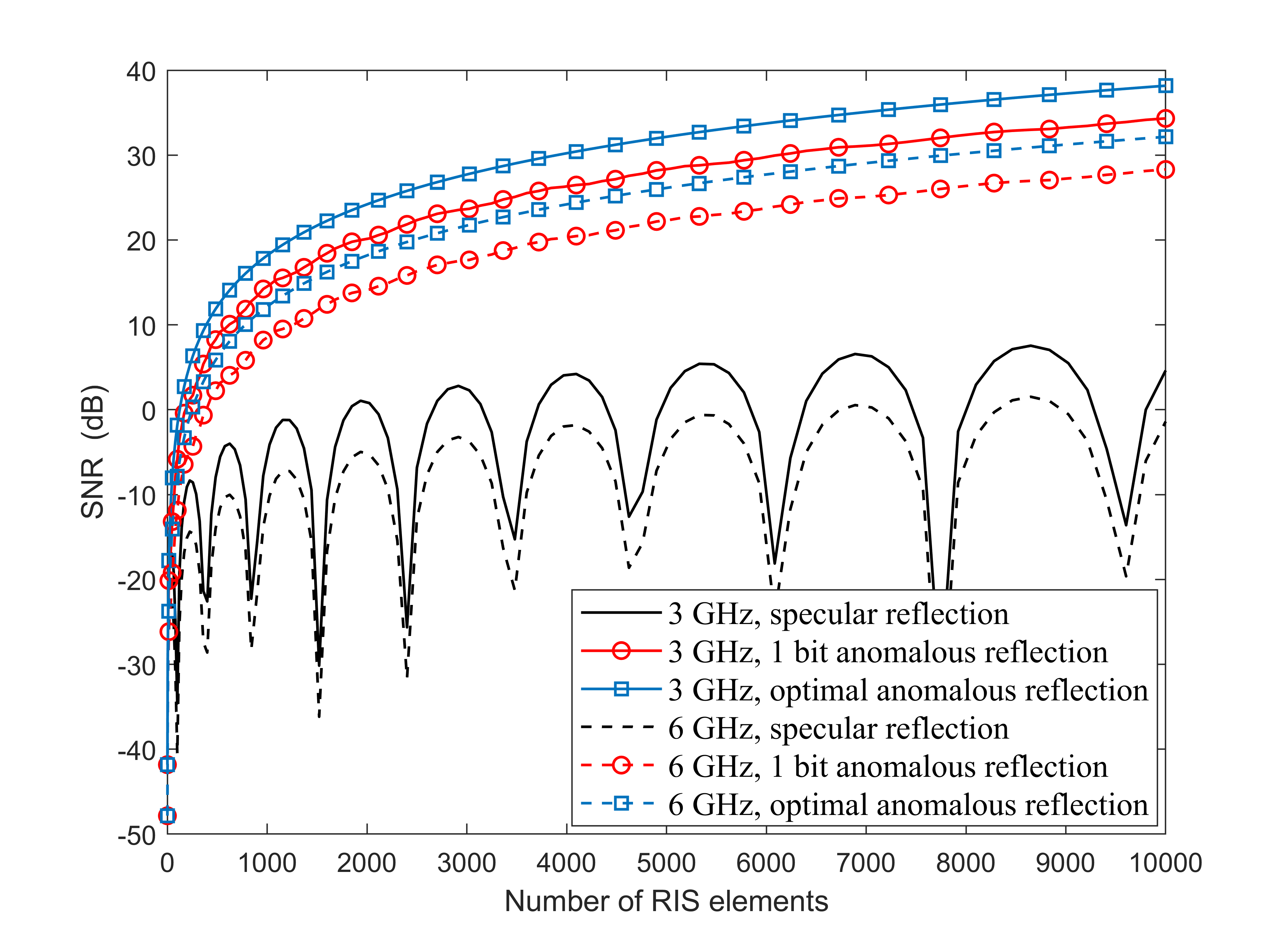}}
\caption{Receiving SNR in RIS cascade link under different RIS configurations.}
\label{fig2}
\end{figure}

In Fig. \ref{fig2}, the SNR at the receiver in the RIS cascade channel is presented for the varying number of RIS elements and the operating regimes.
The coordinates of transmitter, RIS and receiver are $(0,0,10)$, $(-15,15,6)$, and $(-10,30,2)$.
The noise power at the receiver is set to -120 dBm, and the power gain of the transmitter and receiver are both set to 5 dBm.
We have set up 3 and 6 GHz working frequencies, and the number of RIS has been increased from $1\times1$ to $100\times100$. The channels from the transmitter to the RIS and from the RIS to the receiver are both set to the LOS case.
It can be noted that RIS working in a specular reflection regime brings small gains to the received signal. 
Moreover, in the case of non-specular direction, the RIS gain under this regime does not increase steadily with the increase of the number of RIS elements. 
The 1bit and optical anomalous reflection operating regime can effectively improve the receiving power compared to the specular reflection cases.
And with the increase in the number of RIS elements, this gain effect is more obvious. When other conditions remain unchanged, compared with the effect of RIS at different frequencies, we can see that in the lower frequency, the receiving SNR is higher. This is because the design size of the RIS element is related to the working frequency, and the area of the low-frequency RIS element is larger. According to \eqref{AF}-\eqref{V-in-element}, the radiation pattern of the RIS element is positively correlated with the area.

\begin{figure}[htbp]
\centerline{\includegraphics[width=8.5cm]{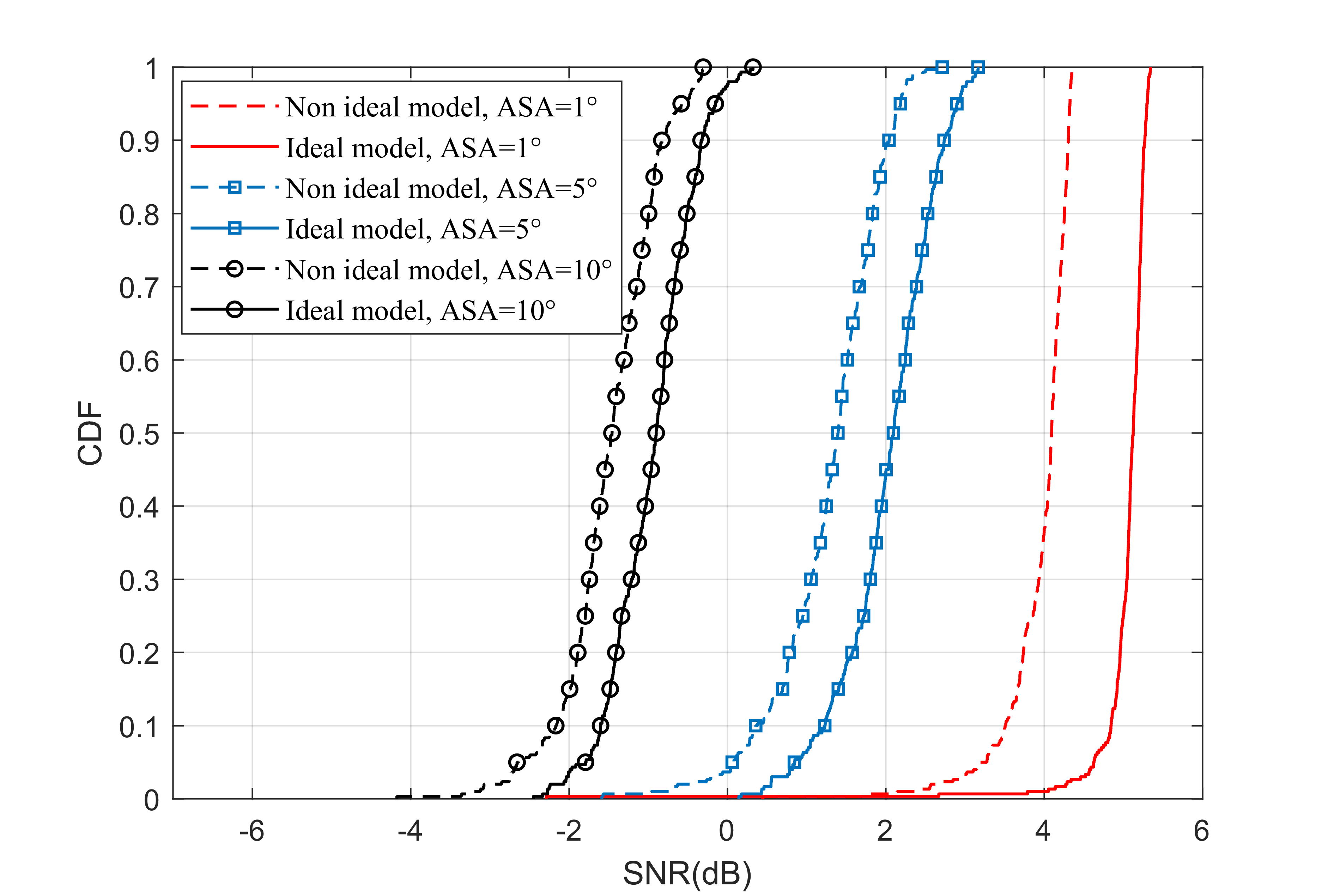}}
\caption{CDF of receiving SNR under different azimuth angle of spread in Tx-RIS channel.}
\label{fig3}
\end{figure}

Fig. \ref{fig3} illustrates the impact of non-ideal phase modulation on the RIS cascade channel under different channel angle spreads. 
Three different ASA are set up in the Tx-RIS channel, which are $10^{\circ}$, $5^{\circ}$, and $1^{\circ}$. The RIS works in an optimal anomalous reflection regime.
It can find that with the increase of ASA in the Tx-RIS channel, the performance of the RIS cascade communication link will be worse.
This is because the greater the angle spread, the lower the energy in the specified incident direction.
In addition, the non-ideal phase modulation brings about 0.5, 0.7, and 1 dB attenuation respectively under the three angle spreads. This means the ideal phase modulation model will overestimate the performance of the cascade channel, and the phenomenon is more obvious when the channel angle spread is small.

\section{Conclusion}

In this letter, a GBSM-based RIS cascade channel model is proposed.
The model can smoothly be compatible with the 3GPP standardization method.
Different from the existing RIS channel model, the proposed model considers the effects of incident path angle on the phase modulation of RIS.
This angle-dependent characteristic is reflected in the radiation pattern of RIS according to the surface equivalence theorem.
The simulation of the RIS radiation pattern verifies the angle and polarization dependence of the proposed model.
Finally, we analyze the impact of non-ideal phase modulation on RIS performance.
The result shows that if the non-ideal phase modulation characteristic is ignored, the performance of RIS will be overestimated.

\bibliographystyle{ieeetr}
\bibliography{reference}

\end{document}